# Aharonov-Bohm effect and coherence length of charge e/4 quasiparticles at 5/2 filling factor measured in multiple small Fabry-Perot interferometers

R.L. Willett, Bell Laboratories, Alcatel-Lucent
L.N. Pfeiffer, K.W. West, Princeton University
M. Manfra, Purdue University

ABSTRACT:

Design of a Fabry-Perot (double point contact) interferometer to measure fractional quantum Hall effect quasiparticle charge properties, and in particular the 5/2 excitations, poses an important trade-off: the device size should be minimized to allow two path interference, since the coherence length of the quasiparticles in the correlated states are expected be limited, yet a small device promotes the dominance of Coulomb charging effects which would overwhelm the Aharonov-Bohm interference effect. In this study a series of small but different size interferometers from the same high density heterostructure wafer are examined for the presence of Coulomb effects versus Aharonov-Bohm (A-B) interference effect when operated in gate configurations that support the 5/2, 7/3, and 8/3 fractional quantum Hall effects. The device sizes vary by more than a factor of three, and over this range explicitly show specific properties of A-B interference, but not Coulomb dominated effects. Given these A-B interference results, the coherence length of the charge e/4 interference is extracted. The coherence length of non-Abelian e/4 quasiparticles is an important parameter for design and development of complex interference devices used to study and apply this exotic excitation. As in prior observations of e/4 excitations, A-B e/4 and e/2 oscillations in alternation are observed in these multiple devices. The amplitudes of the e/4 oscillations are observed to be dramatically reduced for larger area interferometers. Path-length limits are derived from interferometer areas determined directly by A-B measurements, and the attenuation lengths of the e/4 oscillations are shown to be micron to sub-micron scale. This coherence length is consistent with that of the 7/3 excitations measured here, and consistent with theoretical models.

INTRODUCTION:

Experimental study of the 5/2 excitations to examine their statistics [1] has interferometry of the edge currents as the cornerstone of theoretical propositions [2-4]. Realization of that model device is the basis for extensive past measurements [5-7]. The specific type of interferometry device proposed and used to this end is the Fabry-Perot interferometer, comprised of a confined area A that is defined by two sets of constrictions separated by a central region that can be controlled to produce different size areas (see Figure 1). The objective with this device in the 5/2 problem is to produce edge currents that carry the excitations of that state, and with coherent transport around the area A then interfere with a split beam that had partially produced that current. In traversing the area A that current branch will accumulate phase representing not only the Aharonov-Bohm contribution from the encircled magnetic flux, but also a statistical phase by encircling localized excitations in that area. The sine-qua-non of the e/4 excitation as non-Abelian is that for an even number of encircled, localized e/4 quasiparticles the e/4 period Aharonov-Bohm effect oscillations will be expressed, but these oscillations will be suppressed if the encircled localized population is odd. Results consistent with this have been shown in studies that demonstrated e/4 and e/2 period oscillations, with the e/4 expression replaced by smaller amplitude e/2 oscillations in the background occurring aperiodically with interferometer side gate sweep[5-6, 8].

While these results showed data consistent with Aharonov-Bohm effect for integer, 7/3, 5/3, and 5/2 filling factors, the device areas were small and in the size range in which Coulomb effects might be expected. Such Coulomb effects were indeed observed in experiments [9-10] employing lower density samples with markedly different gate configurations that define the interferometer.

These results in total present a problem within the effort to examine correlated state A-B effects: a small size interferometer is pedantically warranted given a smaller device means small path-length and so higher potential for the necessary coherent transport, yet this small size may induce Coulomb effects. These past results [5-6, 9-10] indicate that both A-B and Coulomb effects can occur in interferometers. While the correlated electron state Aharonov-Bohm effect seems to be expressed well in the small devices [5-6], as would be expected due to the smaller path length and so smaller decoherence probability, the small size would suggest that Coulomb effects could be an existent process. The small devices that demonstrated Coulomb effects [9] are substantially different in important parameters; lower electron density and device design. In spite of these points, the possibility of Coulomb contributions in the devices demonstrating A-B interference properties [5-6] at 5/3, 7/3 and 5/2 should be vigorously ruled out.

In this study close examination of the high density systems for potential gross or subtle Coulomb effects is pursued. To this end, a series of devices of different sizes were fabricated using the same heterostructure material as in previous studies [5-7]. With device areas and perimeters ranging over a relative factor of three, interference at filling factors of integers, fractions (7/3), and at 5/2 were studied. Focusing on the interference results from the integer filling factors, the question of whether the oscillations are due to Aharonov-Bohm effect or Coulomb effects is addressed. For all the devices and their various illuminated preparations tested in these high density samples, the oscillations are consistent in detail with the Aharonov-Bohm effect. Two particular A-B properties are shown in these results: first, in all cases, the interference period in B-sweeps does not change over the measureable range of many integral filling factors. Using the same sample that is not subjected to illumination, and so is of lower electron density and lower mobility, a distinctly different property is demonstrated: the oscillations at integer filling increase in period for increasing B-field, as seen in small Coulomb dominated devices [9]. The second effect consistent with A-B properties examines a particular range of integer filling factor oscillations varying both B-field and side gate voltage. Here a distinctive pattern of phase lines consistent with A-B and contrary to Coulomb effects is demonstrated. These results in sum describe that the high density illuminated materials used in these and previous measurements promote the A-B effect and are not conducive to Coulomb dominated processes.

      In the second set of measurements presented here, with the A-B effect affirmed, the coherence length of the e/4 quasiparticle edge propagation is examined. With device areas and perimeters ranging over a relative factor of three, interference at filling factors of integers, fractions (7/3), and at 5/2 were studied to determine the coherence lengths for each. Induced by side-gate sweep, A-B oscillations are observed near 5/2 corresponding to e/4 and e/2 charges in all of these devices, and in patterns of alternation and interchange as seen in previous results [5-6], with the e/4 oscillation amplitudes dramatically reduced for the larger area interferometers. To establish the path-length dependence for each device, interferometer areas were measured by B-field sweep induced A-B oscillations, and limits of perimeter path-length or attenuation length are extracted through several simple models. In the side-gate sweep measurements, diminished e/4 oscillation amplitude with increasing interferometer path-length is shown to be consistent with expected exponential attenuation, as observed for integers in previous studies [15-16]. This attenuation or coherence length for e/4 quasiparticles is micron to sub-micron, in agreement with recent theoretical calculations [8, 11], and is similar to the coherence length of the 7/3 excitations, also measured here. This length is substantially smaller than the IQHE attenuations assessed in these samples, and indicates an important constraint of small interferometer size necessary to observe the e/4 A-B oscillation.

METHODS

At the beginning of this methods section an abbreviated summary of methods and materials is presented. A detailed account of the experiments follows in subsections a-d, reviewing sample and measurement design, sample illumination, gate operation, and data analysis.

Summary: The two-dimensional electron systems employed here are in five different samples cut from the same wafer used in previous studies that exposed e/4 charge A-B oscillations [5-6]. The electron density is roughly $4 \times 10^{11}$ cm$^{-2}$ with mobility of $28 \times 10^6$ cm$^2$/V-sec. As shown in the electron micrographs of Figure 1, the interferometers consist of two nominal quantum point contacts (qpcs), operated as constrictions, with a central gate used to vary the interferometer area via voltage Vs, always of sufficient voltage to induce full depletion of the underlying 2D gas. The qpcs are operated at voltages past full depletion as well, but are of sufficient separation that the longitudinal resistivity $R_L$ always clearly demonstrates the set of fractional quantum Hall states at filling factors 7/3, 8/3, and 5/2. This is shown in the left hand panels of Figure 2. Design of the sample gate structures and the relative dimensions to maintain the overall aspect ratios are shown schematically in Figure 1. The 2DES is about 200nm below the heterostructure surface, with a 40nm layer of amorphous SiN applied to the top of the heterostructure. The aluminum top gates that define the interferometer are evaporated onto this SiN layer.

The interferometer gates are charged to operation values and left at those voltages up to several days before measurements are made. The side gate voltage sweeps are applied at slow rates of roughly 100mV/6hrs, as used in previous studies. Longitudinal resistance $R_L$ and diagonal resistance $R_D$ are measured using low frequency lock-in techniques, and temperatures to roughly 15mK are achieved in two different dilution refrigerators.

Details of the experimental methods are described below.

a) methods and materials

The heterostructure wafer used in these experiments has a relatively high density of $4 \times 10^{11}$ cm$^{-2}$ and high mobility of $28 \times 10^6$ cm$^2$/V-sec, with the 2D electron channel 200nm below the sample surface and is constructed with doping layers both above and below the conducting channel. The top gate structures that define the multiple interferometers used here are shown in Figure 1; as in samples of previous studies [5-7], a 40nm amorphous SiN layer is applied to the sample surfaces and the 100nm thick Al top gates are defined on that surface. The top gate structures are charged to operation values and typically held there for more than 20 hours before B-field sweep measurements are made. Standard lock-in techniques at low frequencies are employed in dilution refrigerators reaching base temperatures of around 15mK. From the labeled contacts a-d in the electron-micrograph of Figure 1, longitudinal resistance $R_L$ is measured by the voltage drop from contacts c to d, with current driven from a to b.

The samples can be briefly illuminated when the gates are not charged to enhance mobility, and so different sample preparations can be achieved as described in prior studies [5-7]. These constitute different sample preparations as the charge localizing potentials change with illumination; zero field resistance from diffuse boundary scattering is observed to be altered significantly for the different illuminations [7]. The coarse differences in transport through the devices of different preparations can also be seen comparing the $R_L$ spectra from filling factors 3 to 2 in the magneto-transport results of Figures 2 to 8.

The two standard top gate designs shown in the electron micrographs in Figure 1 are the basis for the series of interferometers. The interfering edge paths are delineated in the electron-micrograph of Figure 1 with back scattering at the constriction gates marked 1 and 3. The adjusted device dimension parameters x and y are shown in the photo. Lithographically defined device areas in three separate samples are produced using x, y pairs of the following dimensions in μms: device 1; 1.6, 2.0, device 2; 2.1, 3.0, and device 3; 2.8, 3.5, resulting in ratios of areas roughly 3:2:1. The range of functional areas in these devices is also defined by the continuous range of gate voltages applied, and the combined variation of lithographic design and voltages resulted in a range of active device areas from ~0.1 to ~0.5μm$^2$.

In practice, the functional area of the device is substantially smaller than the lithographically defined area since the depletion mechanism is due solely to electrostatic depletion with a large lateral component. An important limitation to this interference measurement is that the top gates must be allowed to equilibrate over long time periods, typically many hours. This limits the effectiveness of swept gate measurements, as intrinsic charge relaxation to changing voltages must be considered.

As in our previous studies of Aharonov-Bohm oscillations near 5/2 filling factor, the value of the side gate voltage change corresponding to e/4 charge is determined by similar measurements at integral filling factors and 7/3 filling factor using the relationship $\Delta V_s \sim \phi/B = h/(e*B)$. With definition of the presence of e/4 interference, the active area of the interferometer is measured by examining B-field sweep measurements over the entire filling factor range near integral filling, with oscillation period in B related to the device area via $\Delta B_1 = \phi/A = h/(eA)$.

To summarize data collection details, the two measurements employed here are; $R_L$ measurement with gate voltage sweep Vs, and measurement of $R_L$ with B sweep. In Vs sweep measurements the gate sweep rate is 100mV/6hrs, with lock-in time constant for $R_L$ set typically at 10 seconds. Vs sweep measurements were performed on these multiple area samples to establish consistency with previous results [5-6]. The results are shown in Figures 9-11, demonstrating the presence of e/4 and e/2 A-B oscillations occurring alternately as previously found. For B sweep measurements, in particular the $\Delta B_1$ measurements in Figure 3-6, the B sweep rate is at or less than 20 Gauss/minute.

b) transport through device: geometrical effects and sample illumination

A crucial part of these studies is achieving longitudinal resistance through devices that demonstrate a confined 2D system able to support the small gapped states at 5/2, 7/3, and 8/3. At the same time, beyond demonstrating these fractions in transport through the interferometers in this range of device areas, the interferometers are able to expose A-B oscillations in resistances $R_L$ and $\Delta R_L$ from sweeping side gate and from sweeping B-field over this range of device dimensions. This indicates that the fundamental properties appear to be consistent over areas differing by a factor of more than three.

As outlined in previous studies [5-7], a layering of amorphous SiN and then deposition of the top gates with no etching is used to preserve the high quality of the device area. To achieve the highest mobilities, these samples require brief illumination before transport measurements, and this illumination is likewise applied to the devices before gate charging. This illumination produces high resolution of the fractional states not only in the adjacent bulk but also in the devices as demonstrated by the results for all device sizes as shown in Figures 2 to 6 and 8. Note the subtle differences in the transport spectrum between preparations of similar coarse area (1, 2, or 3) which can be due to different illumination preparations and/or different gate voltages.

The geometrical arrangement of our contacts and the gross patterning of the top gates further emphasize the importance of these high quality transport traces for these different sample sizes as indications of the preserved correlations within the device area. As shown in the electron micrographs of Figure 1, the number of squares through which transport occurs through the devices ranges approximately from two to more than three. External to the devices the transport contacts to the 2D gas are arranged such that roughly one square of transport area is used. This means that the interferometric device areas are contributing a large proportion of the measured resistance. This geometry emphasizes the contribution from the device area, and high quality of the transport traces through these devices attests to the quality of the confined 2D electron system. This defines an important principal in these measurements: a high quality transport trace through the device ($R_L$) is weighted by the geometrical factor of the device with respect to the bulk. Since $R_L$ is necessarily measured with both bulk and device contributions, a geometry heavily weighted to the bulk (more squares of area there than in the device) can mask a poor device 2D gas quality, showing high quality transport even though the device quality is low.

c) lateral depletion and small active area versus lithographic area

The active area of the interferometers as determined by A-B oscillations is substantially smaller than the lithographic or top gate area, which is consistent with the large distance of the 2D gas from the surface and the large negative voltages applied to the surface gates. With the 2D gas 200nm from the surface, and the additional insulating 40nm SiN layer, at nominal full depletion of the electrons below the gate the border of the active area must be at least 240nm internal from the lithographically defined gate structure. Since the gates are typically operated at much larger negative voltages than full depletion (~ -2V), the active area will be reduced substantially further. Even with this large bias and consequent wide lateral depletion profile, the d.c. transport through the device shows survival of 5/2, 7/3, and 8/3 fractional states: this may be due to the relatively high electron densities used in this study. The density used here is substantially higher than in other studies [9-10]. In our own experiments using samples at their lower densities and without illumination, only a resistive lump is observed between filling factors 2 and 3, with no resolution of fractional states. See Figure 6. Since both higher density and sample illumination are used in our studies, either or both may be the origin of the higher quality transport observed in our samples.

d) data analysis

In the $\Delta R_L$ traces shown the local background resistance has been subtracted to display the oscillations which are typically less than a few percent of the overall resistance. This background is an adjacent average at each data point taken using a large averaging number which allows approximation of a smooth background: the specific number of data points over which the average is taken depends on the density of original data points, which depends on the rate of either the B-field sweep or the side gate change. This smoothed background is then directly subtracted from the raw data.

RESULTS:

The results of two sets of studies are presented here: a) examination of whether the Aharonov-Bohm effect or Coulomb effects are the dominant processes at play in these specific Fabry-Perot interferometers and these specific high density samples; and using the result that the A-B interference process is indeed dominant from this examination, b) extracting an experimentally based estimate of the coherence length of the e/4 quasiparticle propagation in these devices.

a) Aharonov-Bohm versus Coulomb dominated resistance effects

A critical distinction in interferometer measurements is whether oscillations observed are due to the Aharonov-Bohm (A-B) effect or due to Coulomb effects (Coulomb blockade or Coulomb dominated) [12-13]  This is of particular importance when the active area of the device is small: it has been found in previous studies [9] that small area devices display resistance oscillation properties consistent with Coulomb domination.  In larger but similar devices [9-10], oscillations demonstrated A-B properties.

In the measurements of this study, in all the illuminated devices and their respective area ranges acquired through different top gate voltages, the resistance oscillations have shown consistency with A-B oscillations.  This empirical consistency with A-B is demonstrated in the B-field induced oscillations at integer filling factors.  This is shown in the data of Figures 3-5.  Over a large range of B-field, from filling factor 6 to 2, no change in the resistance oscillation period is observed; this holds within different sample preparations of different areas.  This constant B-field period over a large filling factor range is as observed for large devices in the previous studies [9] in A-B, and distinctly different from Coulomb dominated results in smaller area samples of the same study, where B-period is proportional to B.  We see this constant period over a B-field range of more than a factor of three in these samples, and this constant A-B oscillation period implies a constant active area A for the interferometer for each preparation.   This empirical consistency with A-B is observed in all samples tested here, and is distinctly inconsistent with Coulomb domination or Coulomb blockade.

Another property of the presence of the A-B process is shown in interferometer measurement where the side gate is swept to alter the area.  If this measurement is performed at different integer filling factors, it is found [9] that the oscillation period increases linearly with B.  Similar measurements were taken previously [6] in the smaller of the samples of this study, and there this linear increase was observed, consistent with A-B and contrary to Coulomb effects.

The second crucial indication that Aharonov-Bohm oscillations are observed in the samples of this study is close examination of the B-field and gate voltage dependence of the oscillation phase at any given integer filling factor.  It is expected [12-13] that in the case of Aharonov-Bohm oscillation in a plot of increasing B in x, and decreasing side gate voltage in y (less lateral depletion with increasing y) versus resistance in any color code z, the lines of constant phase should have a negative slope.  To the contrary for Coulomb blockade or Coulomb dominated physics the slope of constant phase is positive [13]. Figure 6 is a cursory study of a single integral filling factor over a small range in B-field and side gate voltage, showing the constant phase lines running with negative slope, in agreement with Aharonov-Bohm oscillations. Simple tests of this type on other preparations, measuring phase as a function of B-field and side gate voltage have all yielded negative sloped phase.

The samples used in our studies have significant differences from samples in other studies [9-10, 14], namely density, illumination, no surface etch, and open constrictions to promote the formation of fractional quantum Hall effect correlations. These differences may explain why the smaller areas here present A-B oscillations. The sample densities used here are roughly twice those in the other studies; this higher density promotes both correlation effects and screening capacity, but also a resistance to area change with B-field change. By not etching the sample surfaces, our devices retain their high quality: surface etching as performed on other devices [14] clearly destroys their ability to support correlation effects and also reduces in an uncontrolled fashion the electron density in proximity to the gate structures.

Substantial promotion of correlation effects occurs with sample illumination in our studies, as well as an increase in electron density. Both illumination factors may play a role in supporting A-B versus Coulomb effects. When not illuminated, our samples show poor transport through the device with little evidence for fractional states in the second Landau level. This is displayed in the top panel of Figure 6: this device was not illuminated and can be compared to the same illuminated sample transport of Figure 3. While the illuminated device has a density of $\sim 4.0 \times 10^{11}/cm^2$, the non-illumintated sample density is $\sim 2.4 \times 10^{11}/cm^2$. In the non-illuminated sample, oscillations are observed and shown in Figure 6, panels b to d. The periods of these oscillations are plotted versus their B-field positions in panel e. The periods increase linearly with B-field, in contrast to the constant period observed in the illuminated samples of this study. This linear increase in oscillation period with B-field is as observed in the small devices of previous studies [9] where Coulomb effects are stated to dominate. Illumination is not provided in the other studies [9-10]. This result demonstrates an important empirical difference between the illuminated and non-illuminated devices, respectively showing A-B and Coulomb effects.

b) e/4 excitation coherence length

In this section the following three principal findings are presented: i) observation of e/4 A-B oscillations and their characteristic non-Abelian alternation pattern with e/2 oscillations using side gate sweeps, now observed in three more devices, and with e/4 amplitudes substantially smaller for the larger area devices, ii) A-B effect with B-field sweep observed at integer filling factors in all the various size samples that allows measurement of their area, with consequent assessment of interferometer perimeter length, and iii) from both of these measurements coarse derivation of attenuation lengths for e/4 charge edge propagation, and for comparison e/3 at 7/3 and integer filling.

Figure 8 shows representative transport $R_L$ between filling factors 2 and 3 with A-B oscillations in B-field sweeps for the measured devices. In each device preparation displayed here measurement of the B-swept period at multiple B-field positions shows the same value, consistent with A-B effect, corroborated by side gate voltage sweeps that reveal a linear increase in period at integer filling for 1/B change, as described for the preparations in results section a [6]. The values of the interferometrically active areas A are derived from these swept B-periods according to the relationship $A = \phi/\Delta B = 40$ Gauss-$\mu m^2$ / B-period, where the periodicity reflects the addition of magnetic flux. These periods for the different devices translate to areas ranging from ~0.1 to ~0.4 $\mu m^2$.

As in previous studies on device preparations from this high density heterostructure wafer [5-6], side gate voltage sweeps were performed at fixed B-field resulting in $R_L$ oscillations consistent with A-B oscillations at integral, 7/3, and 5/2 filling factors. Importantly, over the full set of device sizes the oscillations at 5/2 are consistent with charges of e/4 and e/2 as observed before, and the distinctive pattern of alternating dominance of e/4 or e/2 periods is also observed.

Representative data from the three general sized (1, 2, and 3) devices is shown in Figures 9-11. As in previous studies, the side gate sweep voltage $\Delta Vs$ is proportional to the area change, and in the A-B effect the period is inversely proportional to the B-field and edge current charge: $\Delta Vs \sim \Delta A \sim h/(e*B)$. This period can be measured at various filing factors and the charge e* can be determined from this A-B relationship. Figures 9-11 show the gross $\Delta R_L$ data. Also shown are the product of these periods and their respective B-field positions plotted against 1/e*, where for each filling factor the corresponding charge e* is used. If the oscillation periods are consistent with A-B oscillations, the plots should be linear and of the form that $(B)(\Delta Vs)=\alpha/(e*)$. The sets of periods are consistent with A-B effect and the assignments of charge of e/4 and e/2 at 5/2, e/3 at 7/3, and e at integral filling factors. In the $\Delta R_L$ data the resistance oscillations at 5/2 are featured, noting in each the alternation of e/4 and e/2 dominantly apparent oscillations, consistent with non-Abelian e/4 excitations. This alternation is due to the side gate sweep enclosing an alternately odd or even number of localized e/4 quasiparticles as it progresses, which induces expression or suppression of the e/4 oscillations. The e/2 periods are presumed to be due to a persistent background interference by Abelian e/2 quasiparticles that is not parity vulnerable. The e/2 period is observed when the e/4 period is suppressed, with the background e/2 oscillations of generally smaller amplitude than the e/4. These results establish that the e/4 A-B oscillations and their non-Abelian properties can be observed in a range of device areas.

Given that the devices of different areas can display the non-Abelian e/4 interference, these devices can also be used now to deduce the propagation properties of the excitation. For this heterostructure and its 5/2 state, the coherence length of the e/4 excitation propagation can be assessed. To derive this, the amplitudes of the e/4 interference and the respective device active areas must be used.

We return to the B-sweep data in each device which allows extraction of the device area. First, note that these active areas are substantially smaller than the lithographically defined sizes. The resultant small active area in each of the inteferometric devices is likely due to the fact that only lateral confinement via gate bias is used, in conjunction with the relatively large depth of the 2D electron system. A gradual change in density from full depletion under the gates to nominal bulk density near the center of the active interferometric device is induced, with the large gap integer states at filling factors 2 and 3 potentially occupying substantial lateral dimensions. This suggests that the active areas have large aspect ratios, and are 'cigar shaped', with the long dimension along the y direction as marked in Figure 8.

This oblongation of the active area dictates how the perimeter size can be deduced from the area. The two limits to the perimeter, both not practical possibilities, are, at the lower extreme, 4 x (area)$^{1/2}$, and at the upper extreme 2 x ([ qpc separation] + [area/ qpc separation]). In Figure 8 the qpc separation is roughly labeled y. The former does not take into account the necessary length (qpc to qpc dimension) of each device, and the later is not functional as backscattering would be equally likely along the entire length of the device, defining no specific A-B period. A model that accommodates the physical system is accepting a given large aspect ratio of the active area, and deriving the perimeter size from this. If an aspect ratio of 10:1 is used, then the perimeter can be estimated as P = 22 x (A/10)$^{1/2}$. This model has the advantage that it is robust in minor variation of the approximated aspect ratio, as will be shown below, and it roughly simulates what must be the geometrical structure.

A-B oscillations from side-gate sweeps near filling factor 7/3 are shown in Figure 12 for three interferometer sizes; the amplitudes of these oscillations drop as the interferometer area increases. The vertical lines in each mark the period expected at 7/3 from similar measurements at integer filling factors. These 7/3 filling factor amplitudes are plotted versus the device perimeter in the lowest panel of the figure. The perimeter is calculated from the area (derived from the B-sweep A-B oscillations, as described above) using aspect ratios b = l/w = 5 , with active area A=lw, in the expression for perimeter P=2(l+w)= 2w(b+1)=2(b+1)(A/b)$^{1/2}$. The data for 7/3 are plotted in the figure using three different aspect ratios of 5, 10 and 20 to 1. The resultant path-lengths, or attenuation lengths, are ascribed assuming the relationship amplitude ~ exp( -perimeter/$\lambda$), with $\lambda$ the attenuation length. Note that the range of attenuations lengths is from 0.37 to 0.73 μm with the aspect ratios ranging from 5 to 20. This result suggests that the attenuation length is roughly 0.5μm. Also plotted in the lowest panel are the amplitudes of A-B oscillations near integer filling factors: the amplitude does not change substantially over the perimeter range tested here, showing that the attenuation length for the integer edge propagation is substantially longer that that of the 7/3 state, as described in previous studies [15-16].

A-B oscillations near 5/2 filling factor with e/4 period for side gate swept measurements in four devices is shown in Figure 13. As in the 7/3 data, for each trace the vertical lines marking the e/4 oscillations have their periods determined by similar measurements at integer and fractional filling factors (typically 7/3 and 3 or 4). The gate sweep measurements near 5/2 show the characteristic alternation of e/4 period with e/2 period sections for each device size.

The data of Figure 13 show that as the devices become larger the e/4 amplitudes drop substantially. Here the e/4 amplitudes are plotted versus the modeled perimeters for various aspect ratios of the active areas. Again as for 7/3 amplitudes, the results are roughly consistent with the expected picture of A ~ exp-(x/$\lambda$), where $\lambda$ is the attenuation length. For these models assuming a 5, 10, and 20 to one aspect ratios, an attenuation length of ~ 0.5 μm is derived, similar to that for 7/3.

These data provide a rough estimate of the attenuation length of coherent transport of the e/4 excitations; a precise measurement would require a path-length controlled in detail as might be obtained using a Mach-Zehnder interferometer [15-16]. The large lateral depletion of our Fabry-Perot interferometers which supports transmission through the device of the delicate correlated state excitations would presumably still be at play in this alternate format.

      A final consolidation of the model used to explain the side gate sweep data and consistency with non-Abelian e/4 quasiparticles is shown in Figure 14. The schematics describe the particular properties of the model.  For a non-Abelian e/4 quasiparticle charge at 5/2 filling factor [1], this charge is produced in pairs with each charge possessing a Majorana operator, and the pair expressing a degree of freedom in presence or absence of a neutral mode.  This degree of freedom is altered between the two possibilities (presence or absence of the neutral mode) by braiding of another e/4 quasiparticle between the pair. As such, in operation of the interferometer, edge current of charge e/4 circulates around the active area A; if an even number of e/4 are present, the e/4 A-B period is expressed.  If, however, an odd number of e/4 are present, the necessary pair to the odd encircled e/4 is external to the active area A: circulation of the edge e/4 charge then results in a braiding of the internal odd e/4.  This changes the neutral mode status for that pair.  In fact, for each edge circulating e/4 that passes between the in and out e/4s of the pair, the neutral mode status changes.  This produces a disruption in the statistical phase of the e/4 circulation, suppressing the coherent interference of that period.  This suppression of the e/4 interference exposes the background, Abelian e/2 interference.  This results in an alternating pattern of oscillatory periods consistent with e/4 and e/2 charges.  Another experimental example of this alternation, modeled on the side gate sweep progressively changing the enclosed e/4 population, is shown in the lower panel of Figure 14.  In the circumstance of an even number of encircled localized e/4 quasiparticles, the expression of the e/4 period is dependent upon the coherent propagation of the e/4 around the perimeter of area A. From the measurements presented here, in the largest devices tested, this coherence is retained.

CONCLUSION:

The high density, illuminated samples in this study demonstrate Aharonov-Bohm interference over a range of interferometer sizes; A-B properties are expressed in both constant interference periods in B-sweeps over multiple integer filling factors and in the phase measurement side-gate and B field change plots. Only when samples are not illuminated, resulting in lower density and mobility is Coulomb dominant transport expressed. These results emphasize the empirical importance of using high density 2D systems to examine correlation effects in interferometry. It is left as an open issue the limit to this picture: at what smaller device size would these high density heterostructures start to show Coulomb dominated effects, or with lowering density but at the nominally same size would the Coulomb effects enter? Other parameters such as the constriction dimension with respect to the area A, may also play a critical role.

This prominence of interference over Coulomb effects allowed crude assessment of the attenuation lengths for the e/4 edge propagation. The measured attenuation length here for the e/4 quasiparticle propagation is consistent with theoretical studies [11]. Their results present a sub-micron coherence length. The dimension measured here suggests that small interferometric devices are necessary for studying the e/4 quasiparticle propagation. This puts a substantial restriction on the available observation window: smaller devices than those used in this study should present the obstacle of Coulomb blockade that could obscure the interference effects [11], and dimensions larger than those used here present a challengingly small oscillation signal given the exponential extinction. This suggests that heterostructures of higher quality, specifically designed to promote a longer e/4 propagation path, would be needed.

FIGURES CAPTIONS:

Figure 1. Electron micrographs of basic top gate structures defining the interferometers and schematics of the size scaling to achieve multiple area devices. Each interferometer is constructed from two nominal point contacts and central area gates; these gates are Al deposited on an SiN layer on the surface of the heterostructure. Upon full depletion gate pairs labeled 1 and 3 are the nominal quantum point contacts that act to bring the edge currents in proximity and promote back scattering at each, which provides the two interfering paths. Gate pair 2 is independently adjustable and allows for continuous variation of area A. Three different areas in ratios of roughly 1/2/3 are lithographically defined by proportional adjustment of dimensions marked x and y. The point contacts all have a separation of 1μm, and the total areas are changed by either just scaling up the defined internal area or using the structure change showed in the two left electron-micrographs. Further alteration of the areas is achieved using different voltages on the top gates.

Figure 2: three separate sample preparations in the three basic different devices showing transport through the interferometer (left panels), and $R_L$ oscillations (right panels) at integral filling factors. Such oscillations are found in proximity to the integer filling $R_L$ zeroes, and are of the same period for different integer filling in each preparation in each device. The period can be used to derive the area of the interferometer A, if these are Aharonov-Bohm oscillations, using the relationship $\Delta B = \phi/A$, where $\phi = h/e = 40 \text{Gauss-}\mu m^2$ and $\Delta B$ is the period of the oscillations. In this picture the area A (see electron micrograph in Figure 1) is the active area of the interferometer, where one edge current branch transmitted at point contact 1 circles area A and interferes with the other current branch reflected at that point contact. Data are taken at T~20mK.

Figure 3. left panels: top, representative area 1 device magneto-transport through the interferometer showing $R_L$ versus B: note preservation of 5/2, 8/3 and 7/3 longitudinal resistance $R_L$ minima; bottom panel is transport through bulk. Right data panels: oscillations near three different integer filling factors for this preparation of device area 1 where the background $R_L$ has been subtracted to expose the small amplitude oscillations. The oscillation periods do not change for the different integer filling factors, consistent with the Aharonov-Bohm effect as the origin of the oscillations. The gate voltages are unchanged during the B-sweep, and the area is determined by the oscillation period as $A = 40 \text{Gauss-}\mu m^2/\Delta B$. T~25mK.

Figure 4. left panels: top, representative area 2 device magneto-transport through the interferometer showing $R_L$ versus B: again note preservation of 5/2, 8/3 and 7/3 longitudinal resistance $R_L$ minima; bottom panel is transport through bulk. Right data panels: oscillations near three different integer filling factors for this preparation of device area 2. Again, the oscillation periods do not change for the different integer filling factors, consistent with the Aharonov-Bohm effect as the origin of the oscillations. T~25mK.

Figure 5. left panel: Representative area 3 device magneto-transport through the interferometer showing $R_L$ versus B: again note preservation of 5/2, 8/3 and 7/3 longitudinal resistance $R_L$ minima. Right data panels: oscillations near three different integer filling factors for this preparation of device area 3. Again, the oscillation periods do not change for the different filling factors, consistent with the Aharonov-Bohm effect as the origin of the oscillations. In all these illuminated preparations of areas 1, 2, and 3 the primary period at integer filling factor does not change significantly over a large range in magnetic field. This is consistent with the A-B effect and contrary to Coulomb effects (period~B). T~25mK.

Figure 6: top panel: transport through the interferometer, $R_L$ versus B, in another preparation of device 2: this preparation showed no change in oscillation period at different integer filling factors as demonstrated in Figures 3 to 5, consistent with A-B oscillations. Bottom panel: from the same sample preparation, a color plot ($\Delta R_L$ in color) versus magnetic field (x-axis) and side gate voltage (y-axis). The lines of constant phase are generally of negative slope, consistent with A-B effect. The active area of the interferometer in this preparation is ~ 0.21µm$^2$.

Figure 7: Preparation of device area 1 where no sample illumination was applied. The top left panel is the transport through the interferometer: note the lower density of 2.4 x 10$^{11}$/cm$^2$ compared to the illuminated sample (Figure 3), and note the lack of correlation effects between filling factors 2 and 3. Right top panel is transport through bulk, again with no illumination. Middle panels are $\Delta R_L$ versus B near multiple integer filling factors. The periods increase roughly linearly with B, as plotted in the bottom panel: this is consistent with Coulomb effects, and contrary to A-B oscillations.

Figure 8: More transport ($R_L$ versus B), through other interferometer preparations in devices of areas 1, 2, and 3, focusing on the correlation effects between filling factors 2 and 3. Note the variations of these gross transport traces from the transport in Figures 2 to 5, attesting to the differences in device preparation. The right hand electron-micrograph shows a schematic of the active area A, which is substantially smaller than the lithographic area and presumptively oblong, with a large aspect ratio. This guides the modeling of the perimeter estimate from the area measurement.

Figure 9: Sample area 1 gate sweep induced A-B oscillations at 5/2 filling factor. The upper and left hand panels show longitudinal resistance change with side gate sweep at filling factor 5/2 in two different device preparations: the oscillations are consistent with A-B effect at periods for charge e/4. The marked vertical lines of these periods are derived from similar measurements at integers, defining the period corresponding to e/4 charge. Note alternation of e/4 and e/2 periods as observed in side gate voltage excursions in previous studies [5-6]. Linear plot of the period x B versus 1/e* for one preparation, consistent with A-B oscillations is shown in the right hand panel. Temperature in all data is ~ 20mK.

Figure 10: Sample area 2 gate sweep induced A-B oscillations at 5/2 filling factor and at 7/3 filling factors. The top panels show longitudinal resistance change with side gate

sweep at filling factor 5/2 and 7/3 respectively for the same device preparation: again linear plot of the period x B versus 1/e* for this preparation (bottom right panel) is consistent with A-B oscillations. As before the marked vertical lines of the periods in the top panels are derived from similar measurements at integers, defining the periods corresponding to e/4 charge and e/3. Again note alternation of e/4 and e/2 periods for these side gate voltage excursions in the top panel of 5/2 data. Temperature in all data is ~ 20mK.

Figure 11: Sample area 3 gate sweep induced A-B oscillations at 5/2 filling factor and at 7/3 filling factors. As in Figure 10, the top panels show longitudinal resistance change with side gate sweep at filling factor 5/2 and 7/3 respectively for the same device preparation: the linear plot of the period x B versus 1/e* for this preparation is consistent with A-B oscillations. As before the marked vertical lines of the periods in the top and left panels are derived from similar measurements at integers, defining the periods corresponding to e/4 charge and e/3. Again note alternation of e/4 and e/2 periods for these side gate voltage excursion in the top panel of 5/2 data. Temperature is ~ 20mK.

Figure12. Top panels show amplitudes of Aharonov-Bohm oscillations near 7/3 filling factor for change in $R_L$ with $\Delta V_s$ in three device sizes. Bottom panel shows amplitude of these 7/3 interference oscillations versus device perimeter; three different calculations of perimeter are displayed using different aspect ratios of the area A. The respective attenuation lengths λ are shown for each perimeter calculation, where Amplitude~ exp(-perimeter/λ). Attenuation lengths λ are roughly 0.5μm over this range of perimeters. Also shown are amplitudes of integer A-B oscillations for different area devices; the amplitude changes little, consistent with the expected large attenuation length.

Figure13. Top panels: Aharonov-Bohm oscillations near 5/2 filling factor derived using side gate voltage sweeps Vs for four different size interference devices: the smaller area 0.8 is achieved by large negative top gate bias to device area 1. Bottom: 5/2 A-B oscillation amplitude versus device perimeters for these series of devices. Again three different area A aspect ratios are used. The derived attenuation lengths are similar to those displayed at 7/3 filling factor. Temperature is ~25mK.

Figure14. Top panel a: schematic of the e/4 quasiparticle parity in the interferometer that can produce A-B oscillations of e/4 or e/2 period. In the left micrograph/schematic an even number of e/4 quasiparticles are encircled in the interferometer by charges circulating along the edge. In the top right schematic a necessarily paired e/4 quasiparticle set has one e/4 charge within the interferometer and one outside; each contains a Majorana operator. As the charge circulating in the interferometer braids between this separated pair, each passing e/4 charge changes the phase contribution of the pair, eliminating coherence of the interference: this leads to exposure of a presumed background e/2 interference. Lower panel b: alternation of the e/2 and e/4 A-B periods upon gate sweep. The schematic shows the gate excursion progressively changing the enclosed e/4 quasiparticle number in the interferometer, as expected in this model of the data's period alternation. Measurement temperature is 25mK.

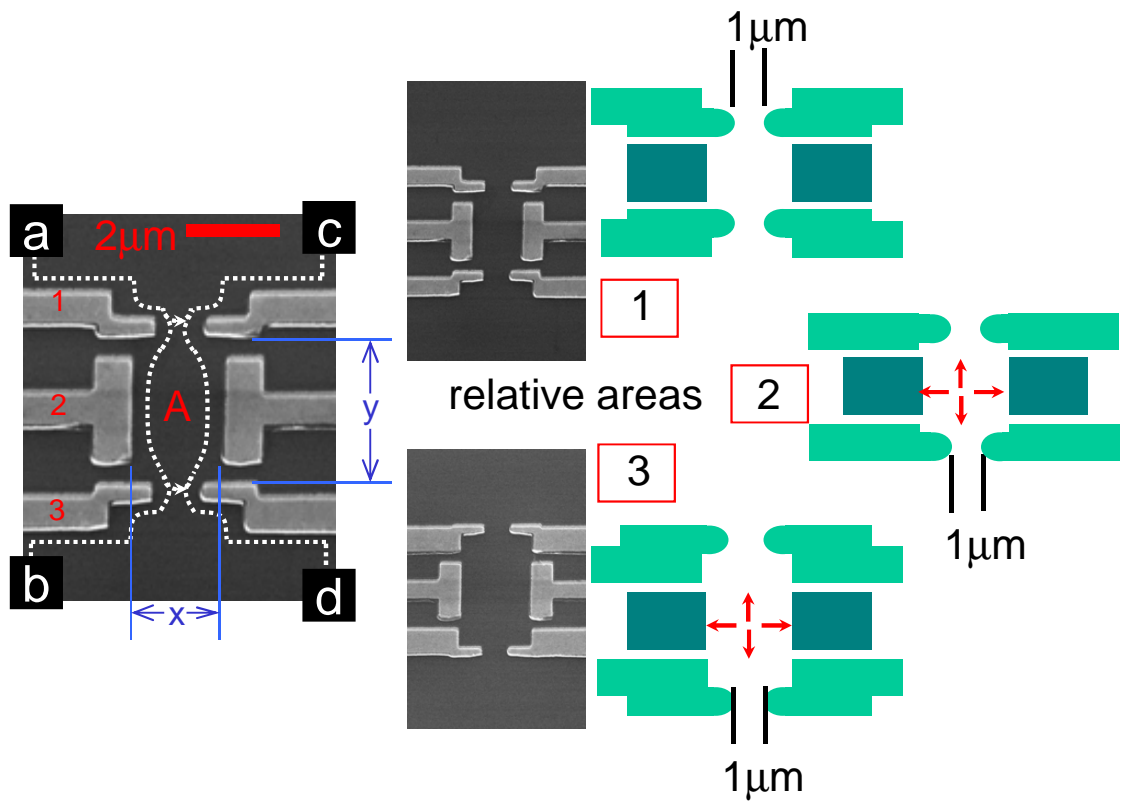

Figure 1

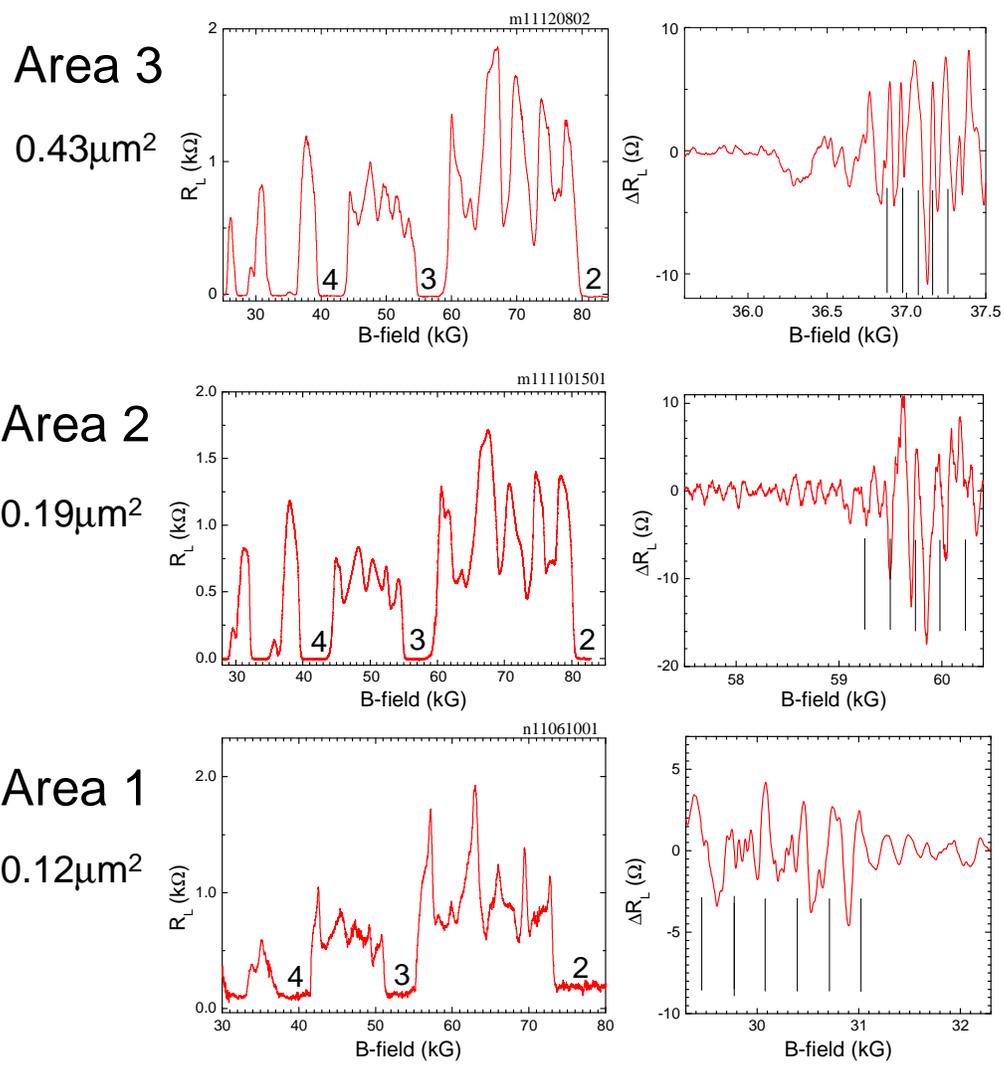

Figure 2

# Figure 3

Area = 0.10μm²

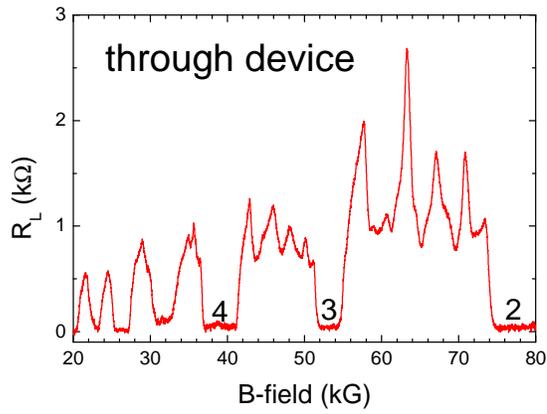

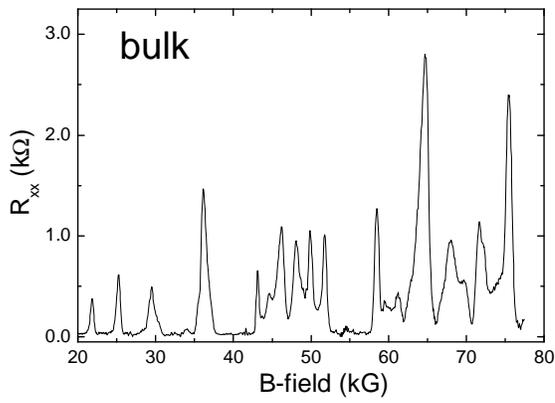

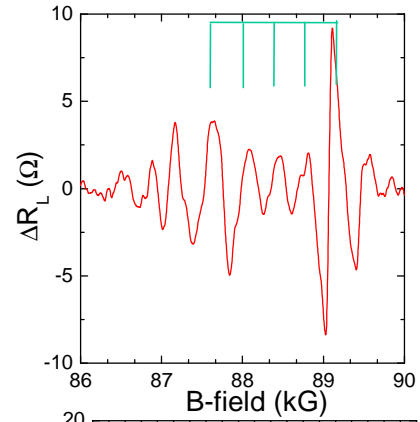

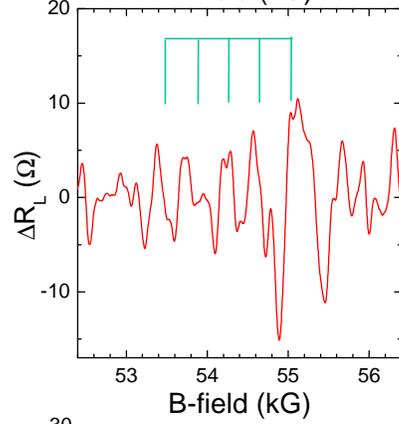

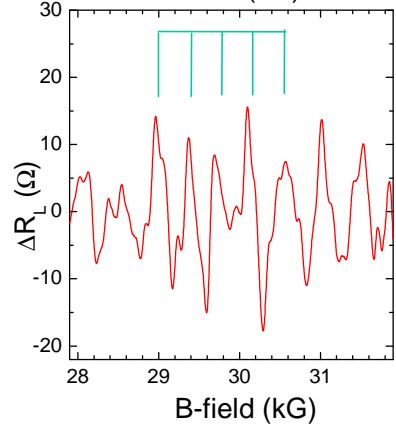

n12062701

# Figure 4

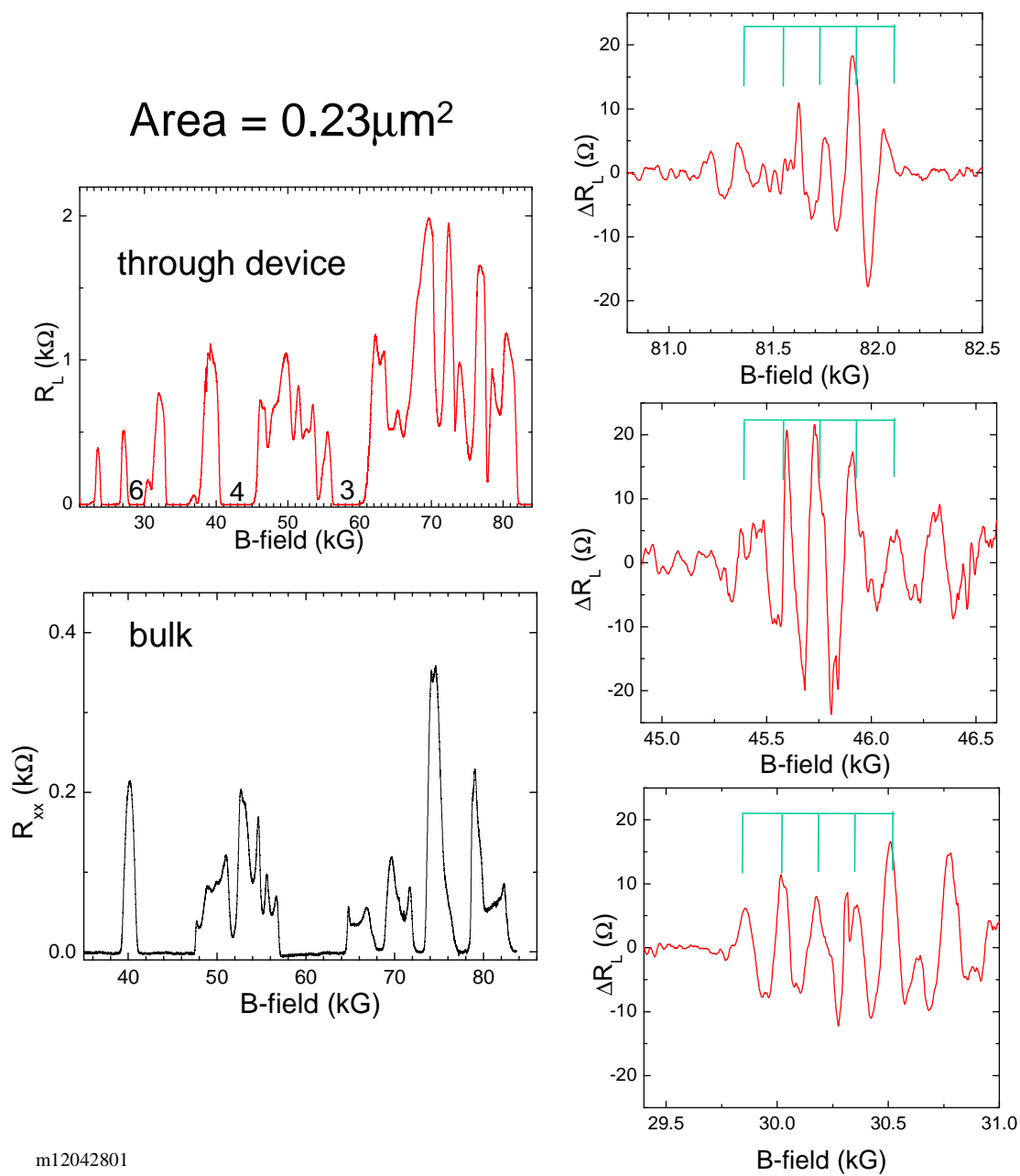

m12042801

# Figure 5

Area = 0.32μm²

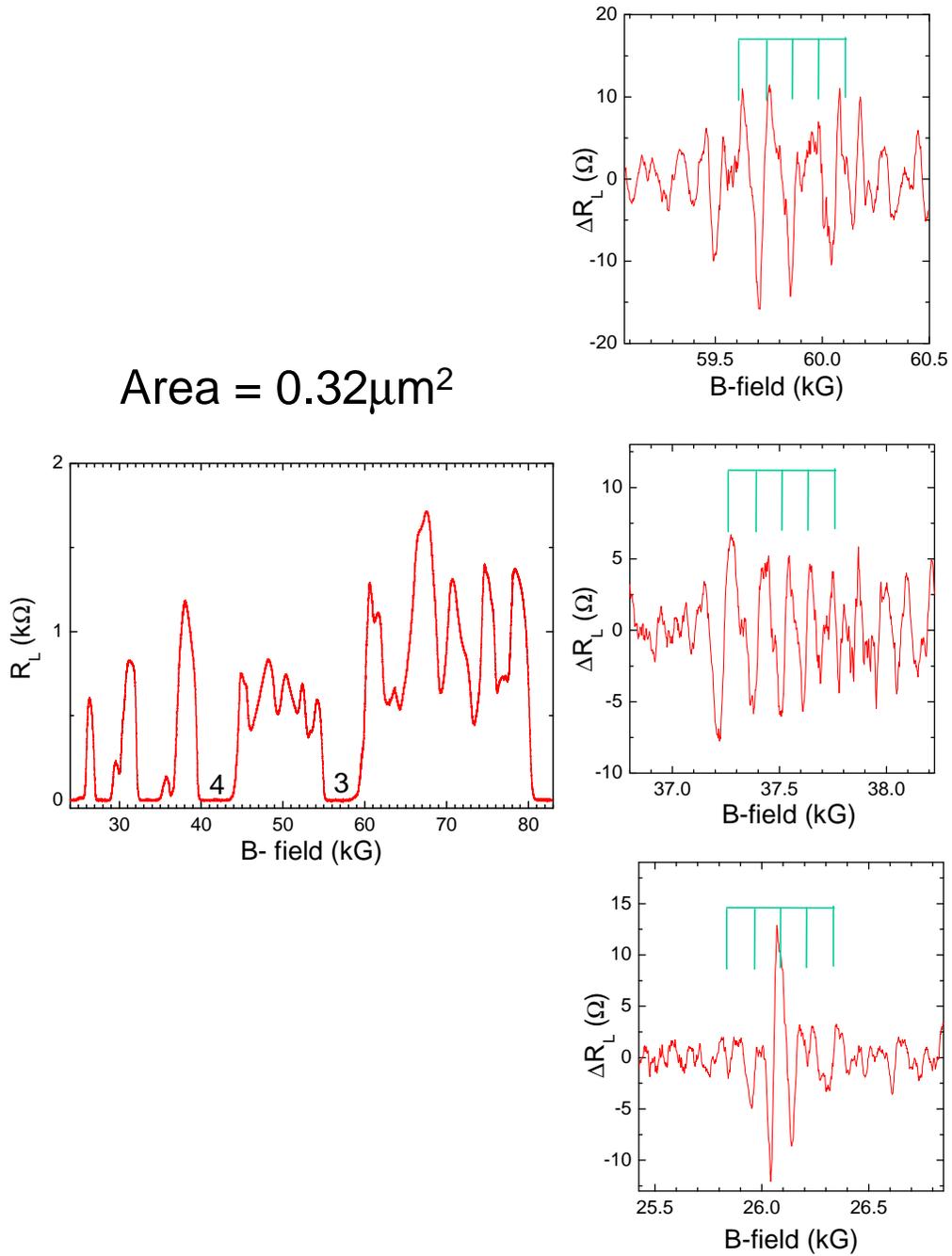

Figure 6

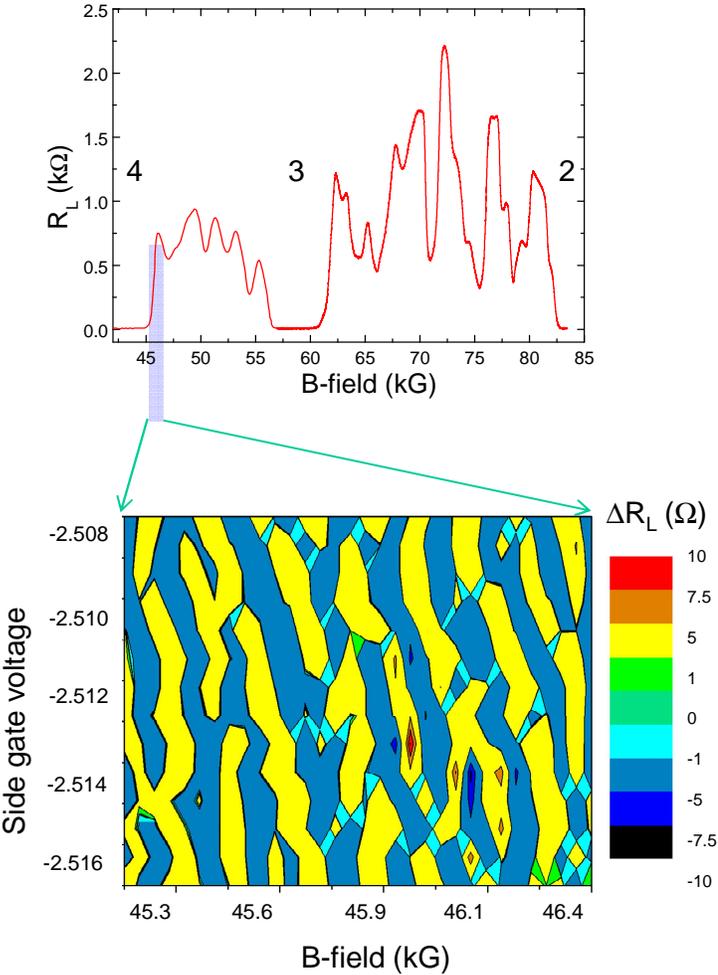

m12061101

Figure 7

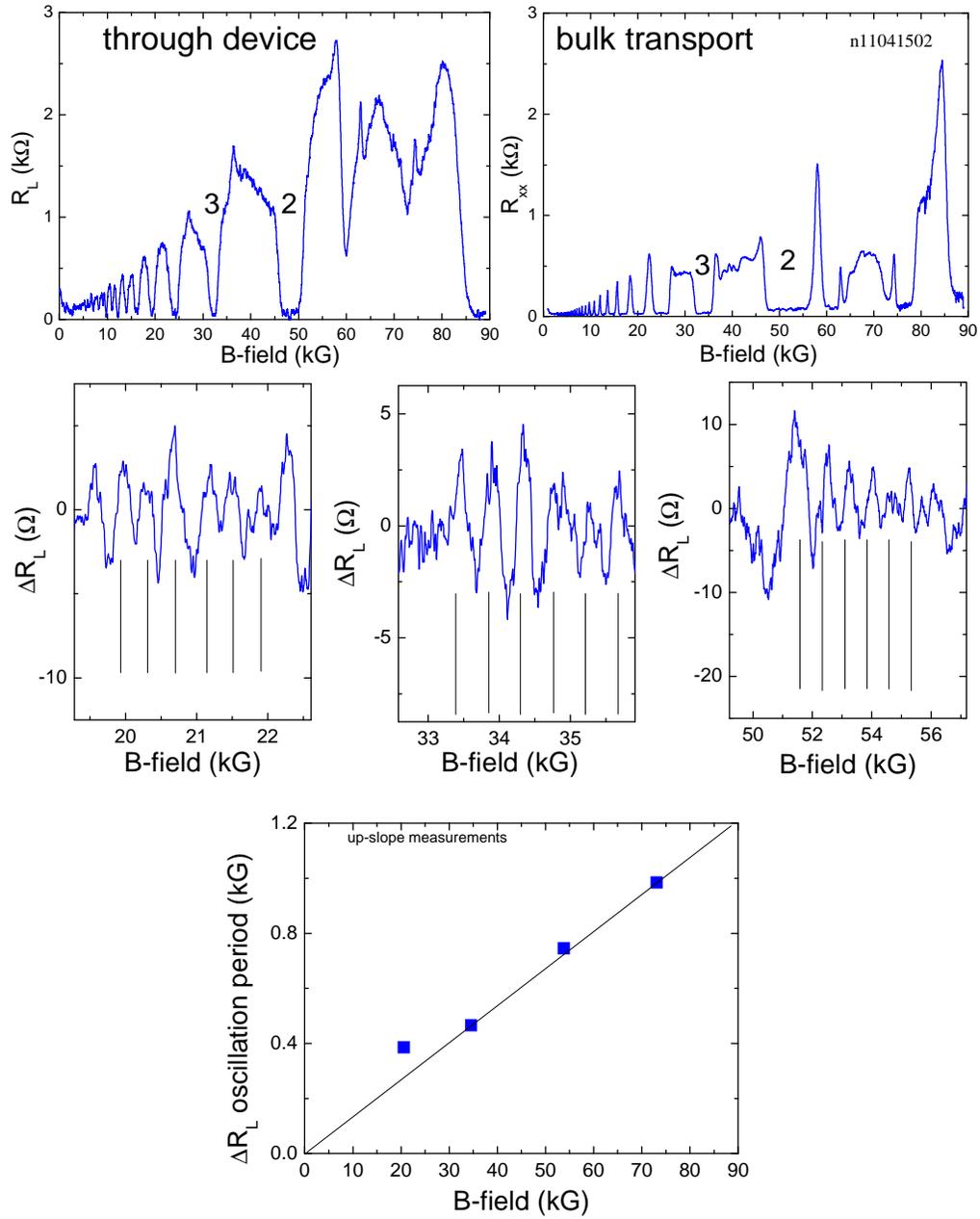

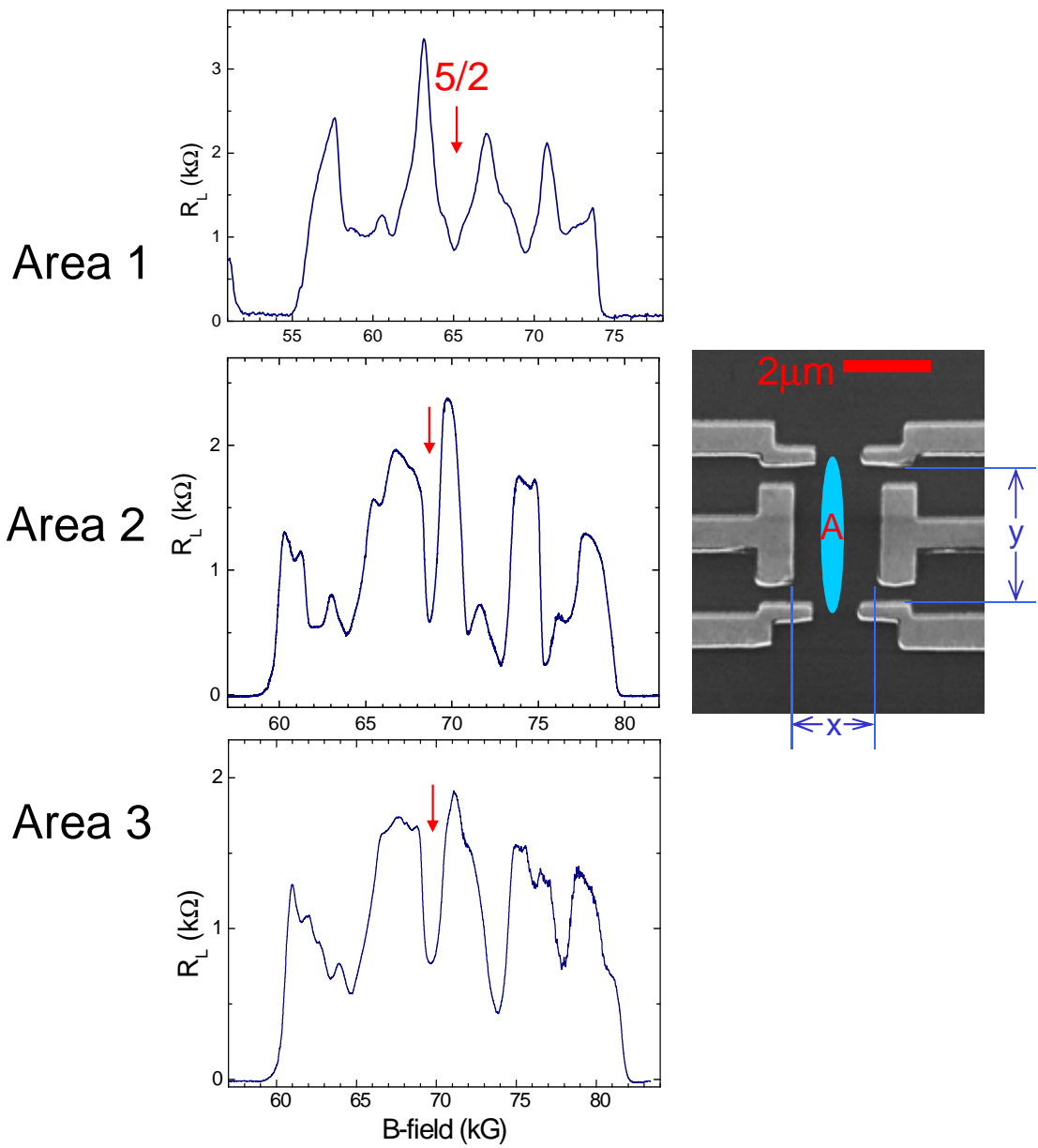

Figure 8

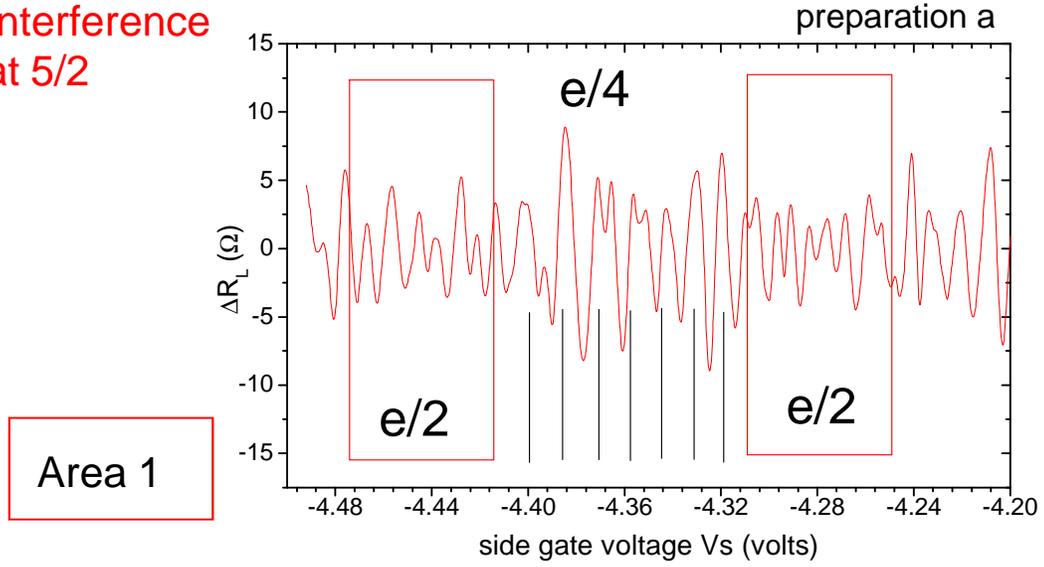
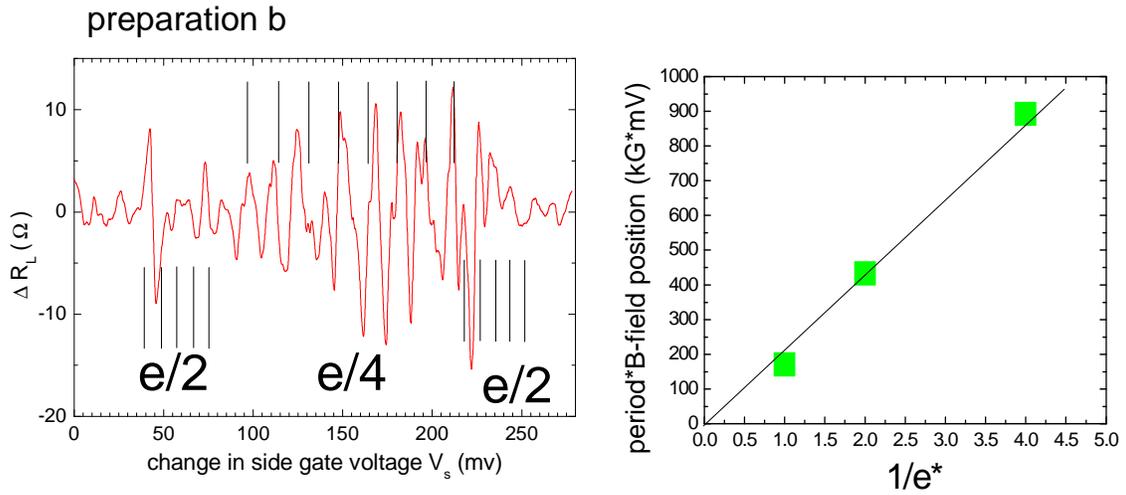

Figure 9

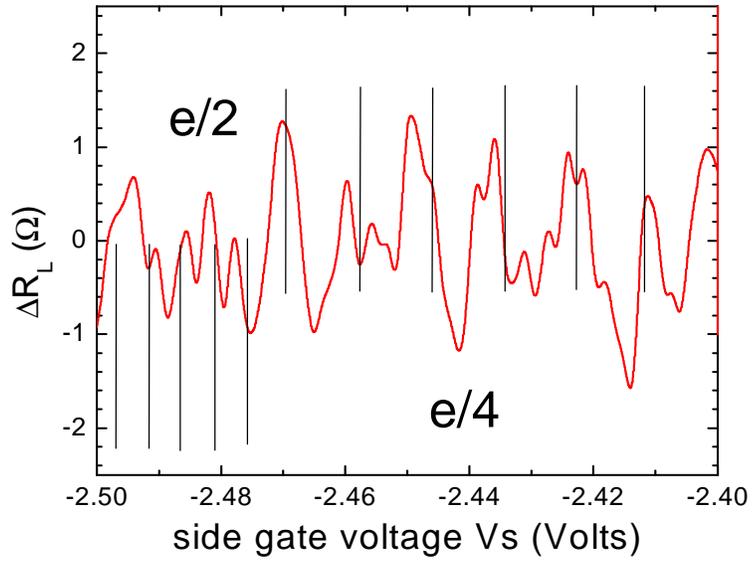

Interference at 5/2

e/2

e/4

Area 2

Near 7/3 filling

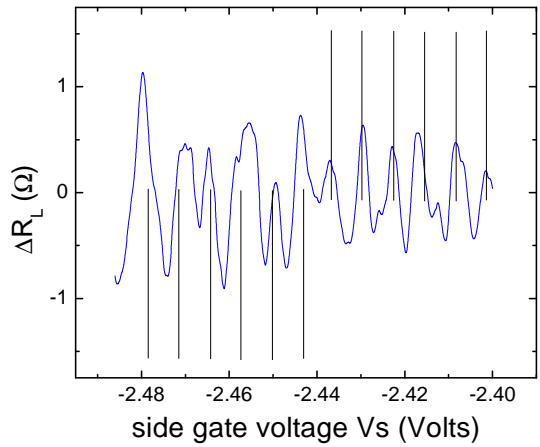
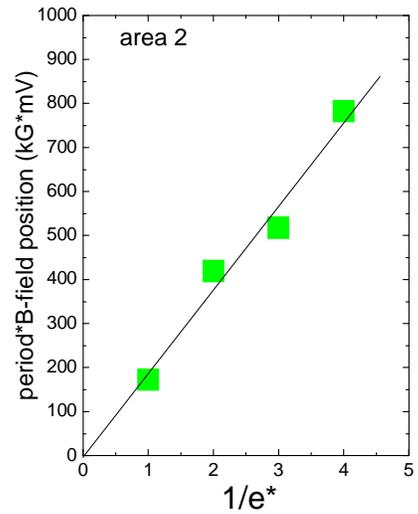

Figure 10

Interference at 5/2

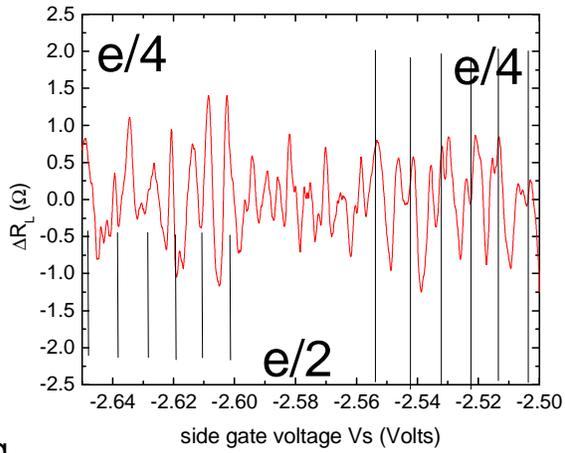

Area 3

Near 7/3 filling

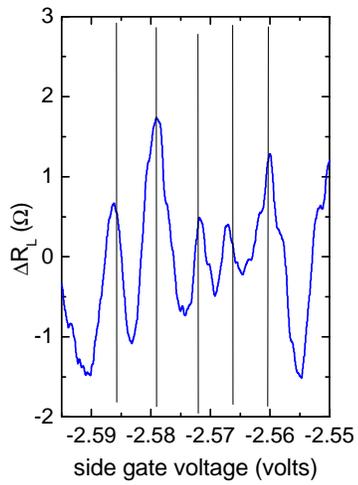

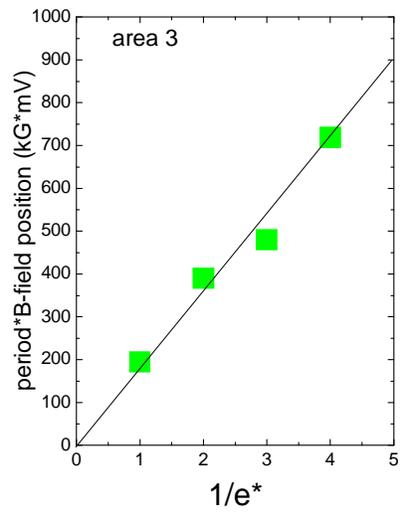

Figure 11

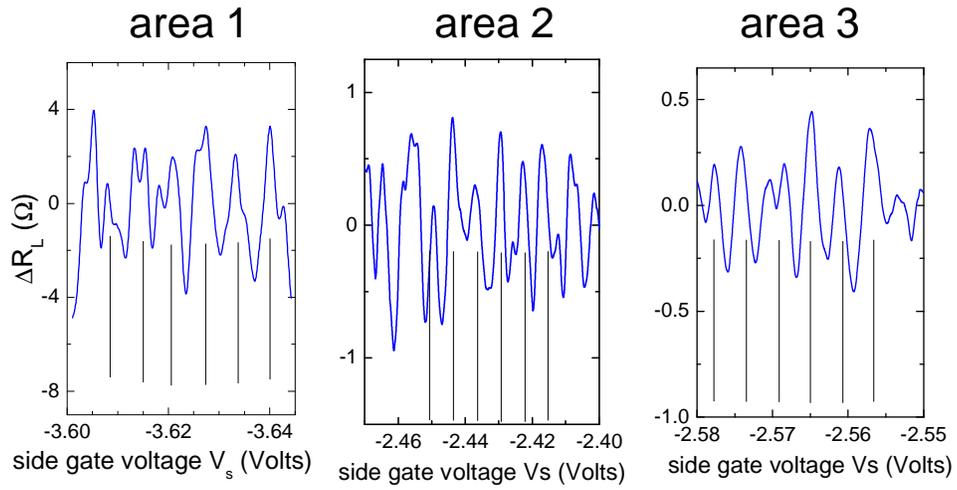
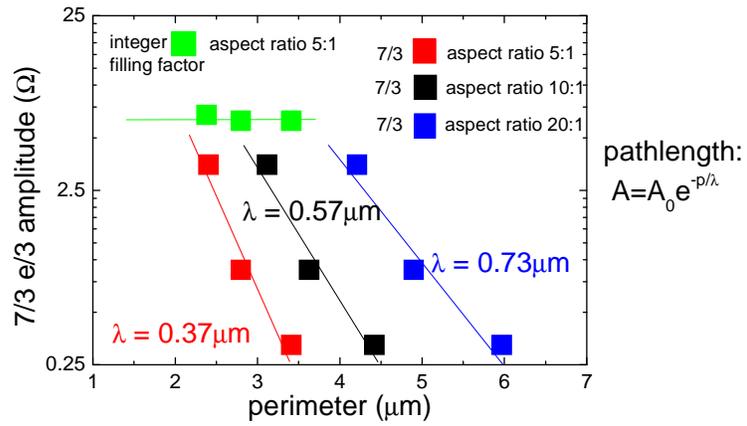

Figure 12

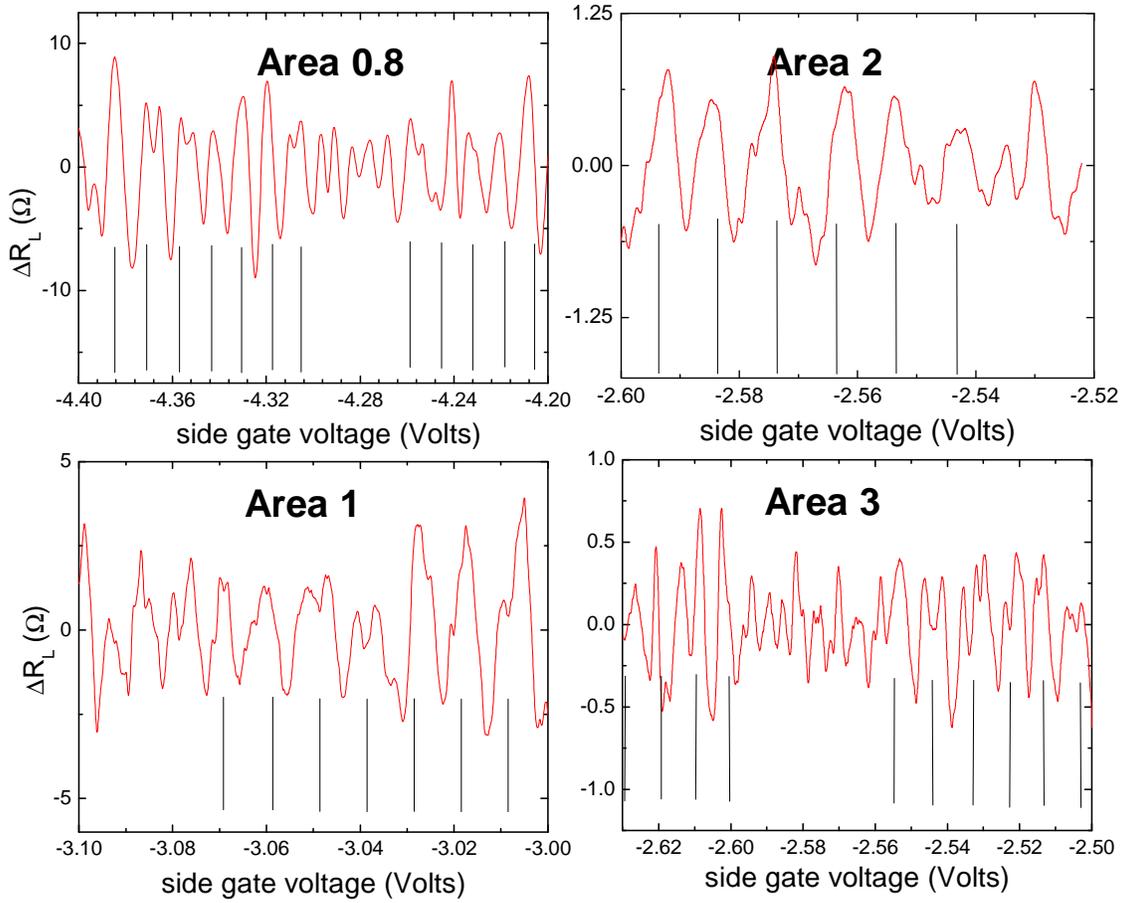

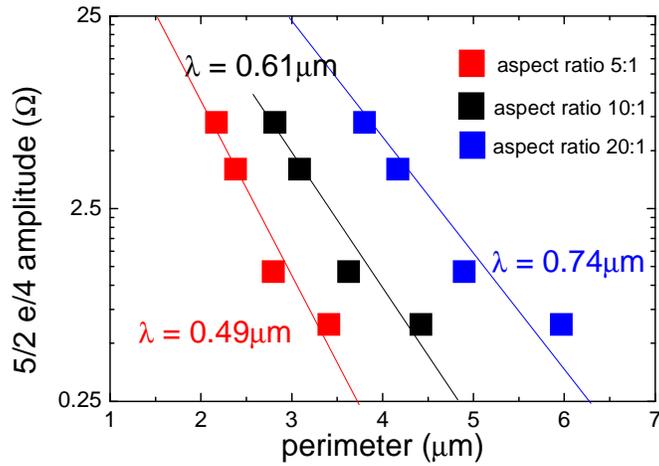

Figure 13

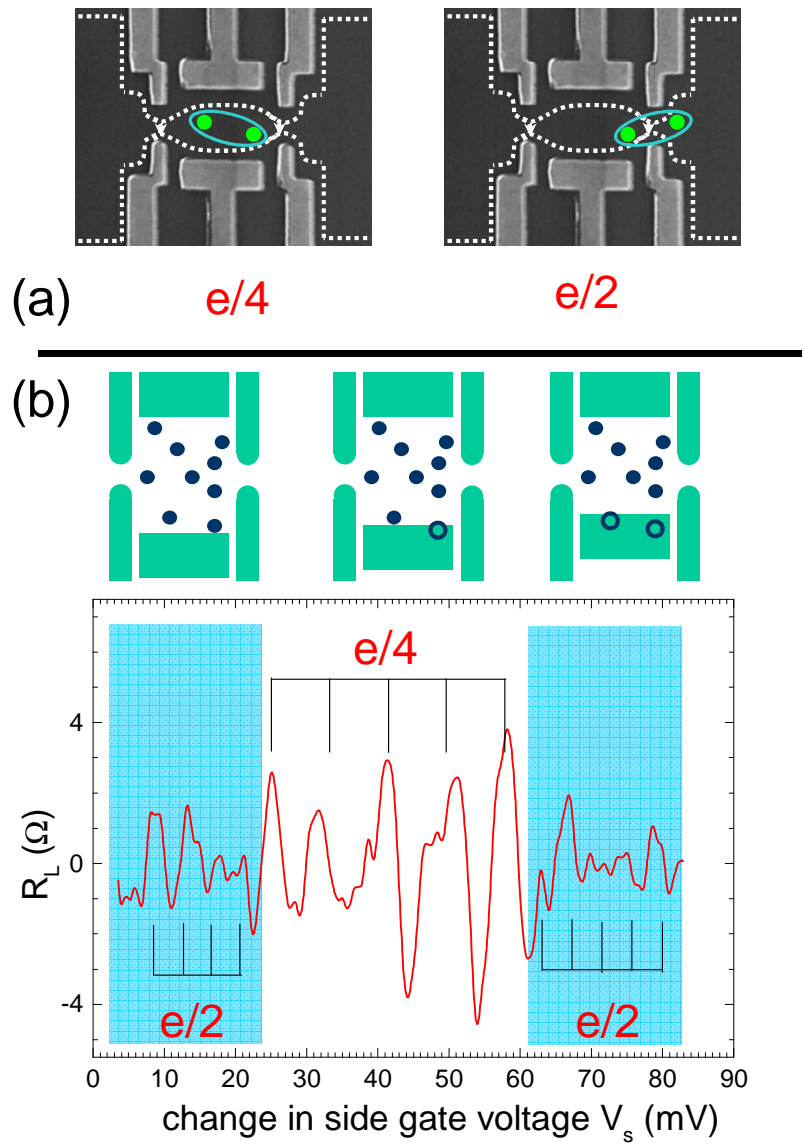

Figure 14